\documentstyle[12pt,psfig]{article}
\newcommand {\cost} {\cos \theta_t}
\newcommand {\eq} [1] {eq.~(\ref{#1})}
\newcommand {\fig} [1] {Fig.~\ref{#1}}
\newcommand {\figz} [2] {Fig.~\ref{#1}#2}
\newcommand {\mst} {m_{\tilde{t}_1} }
\newcommand {\stchic} {\tilde{t}_1 \to c + \tilde{\chi}^0_1}
\newcommand {\stdrei} {\tilde{t}_1 \to W + b + \tilde{\chi}^0_1}
\newcommand {\tanb} {\tan \beta}

\begin{document}
\thispagestyle{empty}
\vspace*{-1 cm}
\hspace*{\fill}  \mbox{UWTHPH-1996-45} \\
\vspace*{1 cm}
\begin{center}
{\Large \bf Higher Order Top Squark Decays
\\ [3 ex] }
{\large W. Porod\footnote{email:porod@merlin.pap.univie.ac.at},
T. W\"ohrmann\footnote{
email: woerman@physik.uni-wuerzburg.de}\\ [2 ex]
${}^1$
Institut f\"ur Theoretische Physik, Universit\"at Wien, A-1090 Wien, Austria\\
${}^2$ Institut f\"ur Theoretische Physik, Universit\"at W\"urzburg,\\
D-97074 W\"urzburg, Germany }
\end{center}
\vfill

{\bf Abstract}

Within the Minimal Supersymmetric Standard Model we study the three body
decay of the lighter top squark $\tilde{t}_1 \to b W \tilde{\chi}^0_1$
and compare this decay with the flavour changing two body decay
$\tilde{t}_1 \to c \tilde{\chi}^0_1$. Here $\tilde{\chi}^0_1$ is the
lightest neutralino which we assume to be the lightest supersymmetric
particle (LSP). We do this for scenarios where two body decays at tree
level are forbidden for the light top squark.
We give
the complete analysis for the three body decay and compare it with the
mentioned two body decay. We discuss our numerical results in view of the
upgraded Tevatron, the LHC and a 500~GeV $e^+ e^-$ Linear Collider.
\vfill
\begin{center}
August 1996
\end{center}
\newpage
\newcommand{\po }{ p_{\tilde \chi_{1}^{0}}}
\newcommand{\pw }{ p_W}
\newcommand{\pbi }{ p_{\tilde b_i }}
\newcommand{\pbj }{ p_{\tilde b_j }}
\newcommand{\pxi }{ p_{\tilde \chi_{i}^{\pm}}}
\newcommand{\pxj }{ p_{\tilde \chi_{j}^{\pm}}}
\newcommand{\pst }{ p_{t}}
\newcommand{\mb }{ m_b}
\newcommand{\mo }{ m_{\tilde \chi_{1}^{0}}}
\newcommand{\mw }{ m_W}
\newcommand{\mbi }{ m_{\tilde b_i }}
\newcommand{\mbj }{ m_{\tilde b_j }}
\newcommand{\mxi }{ m_{\tilde \chi_{i}^{\pm}}}
\newcommand{\mxj }{ m_{\tilde \chi_{j}^{\pm}}}
\newcommand{\mt }{ m_t}
\newcommand{\aiii}{a_{11}^{i \ast}}
\newcommand{\aiij }{a_{11}^{j}}
\newcommand{\aiki }{a_{12}^{i \ast}}
\newcommand{\aikj }{a_{12}^{j}}
\newcommand{\biki }{b_{12}^{i \ast}}
\newcommand{\bikj }{b_{12}^{j}}
\newcommand{\akii }{a_{21}^{i \ast}}
\newcommand{\akij }{a_{21}^{j}}
\newcommand{\bkii }{b_{21}^{i \ast}}
\newcommand{\bkij }{b_{21}^{j}}
\newcommand{\akki }{a_{22}^{i \ast}}
\newcommand{\akkj }{a_{22}^{j}}
\newcommand{\bkki }{b_{22}^{i \ast}}
\newcommand{\bkkj }{b_{22}^{j}}
\newcommand{\ali }{a_{31}}
\newcommand{\alk }{a_{32}}
\newcommand{\bli }{b_{31}}

\section{Introduction}
Supersymmetry is considered as one of the most promising extensions of the
standard model \cite{susy1}. Its search is therefore an important part of the
experimental program at current and
future colliders, namely at the Tevatron, LEP1.5,
LEP2, LHC and a prospective future 500 GeV $e^+ e^-$ Linear Collider.

Within the supersymmetric
extensions of the standard model the minimal supersymmetric
standard model (MSSM) \cite{susy2,susy2a} is the most investigated one.
It contains beside the known SM-particles spin $1/2$ partners for the
gauge bosons (bino,wino,zino,gluino), five physical Higgs bosons (two
scalar $h^0,H^0$, one pseudoscalar $A^0$ and two charged $H^{\pm}$)
and their spin $1/2$ partners (higgsinos). The $ SU(2) \times U(1)$
interaction eigenstates bino, zino, wino and higgsinos mix leading to
mass eigenstates called neutralinos $\tilde{\chi}^0_{i}$ (i=1,2,3,4)
and charginos
$\tilde{\chi}^{\pm}_{j}$ (j=1,2) (see e.g.~\cite{Bartl89} and references
therein). Each fermion has two spin zero partners called sfermions
$\tilde f_L$
and $\tilde f_R$, one for each chirality eigenstate: the mixing
between $\tilde f_L$ and $\tilde f_R$ is proportional to the corresponding
{\it fermion} mass, and so negligible except for the third generation.

The main parameters for the following discussion are $M '$, $M_2$,
$\mu$, $\tan \beta$, $M_{D_i}$, $M_{Q_i}$, $M_{U_i}$, $A_{d_i}$ and $A_{u_i}$.
$M'$ ($M_2$) is the $U(1)$ ($SU(2)$) gaugino mass. In the following we
will assume the GUT relation $M' = 5/3 \tan^2 \theta_W M_2$.
$\mu$ is the parameter of the higgs potential and $\tan \beta = v_2 / v_1$
where $v_i$ denotes the vacuum expection value of the Higgs doublet $H_i$.
$M_{D_i}$, $M_{Q_i}$ and $M_{U_i}$ are soft SUSY breaking masses
for the squarks, $A_{d_i}$ and $A_{u_i}$ are trilinear Higgs-squark couplings.

The mass matrix for squarks in the $(\tilde{f}_L,\:\tilde{f}_R)$ basis
has the following form:

\begin{eqnarray} \label{sqm}
  {\cal M}^2_{\tilde{f}_i} = \left(\begin{array}{ll}
                       m_{\tilde{f}_{Li}}^{2} & a_{f_i} m_{f_i} \\
                       a_{f_i} m_{f_i}  & m_{\tilde{f}_{Ri}}^{2}
                   \end{array} \right)
\end{eqnarray}
with
\begin{eqnarray}
  m_{\tilde{u}_{Li}}^{2}&=& M_{Q_i}^2 + m_{u_i}^2 + m_Z^2\cos 2\beta\,
  (\textstyle \frac{1}{2} - \frac{2}{3} \sin^2\theta_W ) , \nonumber \\
   m_{\tilde{u}_{Ri}}^{2}&=&M_{U_i}^2 + m_{u_i}^2 + \textstyle \frac{2}{3}
                        m_Z^2\cos 2\beta\,\sin^2\theta_W , \nonumber \\
  m_{\tilde{d}_{Li}}^{2}&=&M_{Q_i}^2 + m_{d_i}^2 - m_Z^2\cos 2\beta\,
    (\textstyle \frac{1}{2} - \frac{1}{3} \sin^2\theta_W ) , \nonumber \\
  m_{\tilde{d}_{Ri}}^{2}&=&M_{D_i}^2 + m_{d_i}^2 -
      \textstyle \frac{1}{3} m_Z^2\cos 2\beta\,\sin^2\theta_W ,
\label{smatrix}
\end{eqnarray}
and
\begin{eqnarray}
  a_{u_i}m_{u_i} = m_{u_i}(A_{u_i} - \mu\cot\beta ), \hspace{4mm}
  a_{d_i}m_{d_i} = m_{d_i}(A_{d_i} - \mu\,\tan\beta )
  \label{offdiag}
\end{eqnarray}
where $i$ is a generation index ($ u_i= u,c,t; \, d_i= d,s,b$)
which will be suppressed in the following.

The mass eigenstates $\tilde{f}_1$ and $\tilde{f}_2$ are related to
$\tilde{f}_L$ and $\tilde{f}_R$ by:
\begin{eqnarray}
  {\tilde{f}_1 \choose \tilde{f}_2} =
    \left(\begin{array}{ll}
        \cos \theta_{f} & \sin \theta_{f} \\
       -\sin \theta_{f} & \cos \theta_{f} \end{array} \right)\:
    {\tilde{f}_L \choose \tilde{f}_R}
  \label{mixing}
\end{eqnarray}
with the eigenvalues
\begin{eqnarray}
  m_{\tilde{f}_{1,2}}^2 = \textstyle \frac{1}{2}\,
       (m_{\tilde{f}_L}^2 + m_{\tilde{f}_R}^2) \mp
         \frac{1}{2} \sqrt{(m_{\tilde{f}_L}^2 - m_{\tilde{f}_R}^2)^2
                           + 4\,a_f^2 m_f^2}.
  \label{eq:masses}
\end{eqnarray}
The mixing angle $\theta_{f}$ is given by
\begin{eqnarray}
  \cos \theta_{f} = \frac{- a_f m_f}
           { \sqrt{(m_{\tilde{f}_L}^2-m_{\tilde{f}_1}^2)^2 + a_f^2 m_f^2}},
  \hspace{5mm}
  \sin \theta_{f} = \sqrt{ \frac{(m_{\tilde{f}_L}^2-m_{\tilde{f}_1}^2)^2}
                   {(m_{\tilde{f}_L}^2-m_{\tilde{f}_1}^2)^2 + a_f^2 m_f^2}}.
 \label{eq:mixangl}
\end{eqnarray}
Analogous expressions are also valid for sleptons.

Due to the fact that the off-diagonal terms in \eq{sqm} and therefore
$\cos \theta_f$ is proportional to the fermion mass
the mixing can safely be neglected
for the first two gernerations but in general not for the third generation.
In particular one expects for the top squarks due to the
huge top mass \cite{Ellis83} a strong mixing and a possible big mass splitting
with one light top squark. In the following the top squark (bottom
squark) will be denoted by stop (sbottom).

In general sfermions decay according to $\tilde{f}_k \to
\tilde{\chi}^0_i \, f$ and  $\tilde{f}_k \to\tilde{\chi}^{\pm}_j \, f'$,
where we assume as usual that the $
\tilde \chi_1^0 $ is the lightest supersymmetric particle (LSP).
Contrary to the other sfermions, where the flavour conserving
decay into the lightest
neutralino is always possible, the decay of the stop into the lightest
neutralino will be kinematically forbidden for stop masses accessible at the
Tevatron. Therefore the phenomenological
analysis of stop signals is different from those of other squarks.
Due to the big difference between
the top mass and the bottom mass, even in many scenarios, where the decay
$\tilde t_1 \rightarrow t \tilde \chi_{1}^{0} $ is kinematically forbidden,
the decay
into the b-quark and the lighter chargino, which is heavier than $\tilde
\chi_{1}^{0}$, could still be possible. Since the lower mass limit of
$\tilde\chi_{1}^{+}$ is about 65 GeV \cite{LEPSEARCH} even for
light stops accessible at the current working Tevatron, this decay mode
cannot be excluded. The decay $\tilde{t}_1 \to b \tilde{\chi}^+_1$
will obviously dominate over higher order decays if it is kinematically
allowed.

In the case that this mode is kinematically forbidden,
we have to consider higher order decays
either at loop level or into more than two particles.
There are two competitive modes for a stop accessible at LEP1.5/2 or
the current Tevatron. One possibility is the flavor
changing two body decay $\tilde t_1 \rightarrow c \tilde \chi_{1}^{0}$
occuring at one loop level.
The other possibility is the four body decay into a b-quark, the LSP and
two fermions. In \cite{hikasa} it has been
shown, that for each choice of parameters the one loop decay will be
the dominating one.

For the current working Tevatron both scenarios
($m_{\tilde{t}_1} > m_{\tilde{\chi}^+_1} + m_b$ or
$m_{\tilde{t}_1} <  m_{\tilde{\chi}^+_1} + m_b$)
were considered in recent investigations \cite{tata3}.
Since for $m_{\tilde t_1 } =m_t$ the cross section for the
stop will be smaller by one order of magnitude than that for the top,
$\sigma (\tilde t_1 \bar{\tilde t_1 } ) \stackrel{<}{\sim }
\frac{1}{10} \sigma (t \bar t)$,
this investigation for the current working Tevatron was done for stops
lighter than 120~GeV.
It was figured out, that in both cases the standard model background would be
reducible by appropriate cuts and the stop signal should be distinguishable
from comparable SM processes. Since the stop was not discovered by the
Tevatron, new bounds on the masses of the lighter stop and on the LSP were
found \cite{D0} assuming that the stop decays into a $c$ quark and the LSP.

With the upgraded Tevatron, the LHC or a 500 GeV Linear Collider an
enlarged stop mass range will be  accessible. Due to the
structure of the neutralino and chargino mass matrices,
$m_{\tilde \chi_{1}^{+}} \stackrel{<}{\sim}
2 m_{\tilde \chi_{1}^{0}}$ holds. Therefore we can choose
in the mass range beyond
160 GeV parameters, where $m_W + m_{\tilde \chi_{1}^{0}}+m_b < m_{\tilde t_1}
< m_{\tilde \chi_{1}^{+}} +m_b $, leading to scenarios,
where the three body decay $\tilde t_1 \rightarrow b W \tilde \chi_{1}^{0} $
is kinematically allowed but the two body decay $\tilde t_1 \rightarrow b
\tilde \chi_{1}^{+}$ is still forbidden.
The mass range $m_{\tilde t_1} >160$~GeV will be accessible at the
mentioned future colliders.
It is, therefore, important to study how the rate for the three body decay
compares to that for the flavour changing two body decay
$\tilde t_1 \rightarrow c \tilde \chi_{1}^{0}$ \cite{susy2a}.

In this paper we will study the physics of a stop, which is too heavy
to be probed at colliders currently in operation
but accessible for the upgraded Tevatron, the LHC
or a 500 GeV Linear Collider in scenarios, where
the two body decay into the b quark and the lighter chargino is forbidden.
We will give a full analysis of the three
body decay $\tilde t_1 \rightarrow b W \tilde \chi_{1}^{0} $ and will compare
this decay with the flavor changing two body decay $\tilde t_1 \rightarrow
c \tilde \chi_{1}^{0} $.

This paper is organized as follows. In Section 2 we give the analytical
expressions for the invariant amplitudes of the decays considered here together
with the respective parts of the Langrangian of the MSSM.
In Section 3 we discuss the
total width of the three body decay for scenarios accessible either at an
upgraded Tevatron, the LHC
or at a 500 GeV Linear Collider and compare it with
the flavor changing one.
We conclude in Section 4 with some general remarks . The squared matrix element
is given in the Appendix.

\section{Analytical Calculation of the Widths }

In this Section we will describe the analytical calculation of the decays
considered in this paper. The explicit expressions for the squared amplitudes
are listed in the Appendix.

We first describe briefly those parts of the MSSM, which we will use for our
calculations following the notation of \cite{Bartl89,tata1}. The Feynman graphs
for this process are shown in \fig{graph}.

The neutralino sfermion fermion couplings and the chargino sfermion fermion
couplings used here, we get from the respective parts of the Langrangian,
\begin{eqnarray*}
{\cal L}_{f \tilde f \tilde \chi_{i}^{0}} & = & g
\sum_{k=1}^2 \left[ \bar f
\left( b^{f}_{ki} P_L + a^{f}_{ki} P_R \right)
{\tilde \chi_{i}^{0}} \tilde f_{k} + h.c. \right] \\
{\cal L}_{b \tilde t_1 \tilde \chi_{i}^{\pm }} &=& g \tilde t_{1} \bar b
\left( l^{t}_{1i} P_R + k^{t}_{1i} P_L \right)
{\tilde \chi_{i}^{ - }} + h.c.
\end{eqnarray*}
The neutralino-chargino-W coupling entering in graph 2 we get from
\[
{\cal L}_{ W \tilde \chi_{i}^{\pm } \tilde \chi_{j}^{0}} = - g W_{\mu }^{-}
\bar {\tilde \chi_{i}^{ - }} \left[  {\cal O}^{L}_{ji} P_L +
{\cal O}^{R}_{ji} P_R \right] \gamma^{\mu } \tilde \chi_{j}^{0}  + h.c.
\]
where $P_{R,L}= \frac{1 \pm \gamma_5}{2} $.\\
Finally we get the stop-sbottom-W coupling from
\[
{\cal L}_{\tilde t_1 \tilde b_j W} = \frac{-ig}{\sqrt{2}} \left\{
\cos \theta_b \cos \theta_t \tilde b_{1}^{\dagger }
\stackrel{\leftrightarrow }{ \partial_{\mu }} \tilde t_{1} -
\sin \theta_b \cos \theta_t \tilde b_{2}^{\dagger }
\stackrel{\leftrightarrow }{ \partial_{\mu }} \tilde t_{1}\right\}
W_{\mu }^{+} + h.c.
\]

The invariant amplitudes for the decay width
\footnote{For some subtilities concerning the fermion flow we refer to
\cite{denner}} are given by
\begin{eqnarray} \label{m1}
{\cal M}_{1,i=1,2} & = & -\frac{g^2}{\sqrt{2}} f_{i}(b)  \cos \theta_t
\frac{(p_{\tilde t} + \pbi )^{\mu }}
     {\pbi^{2} - \mbi^{2} - i \mbi \Gamma_{\tilde{b}_i} } \bar u (p_b )
\left[ b_{i1}^{b} P_L + a_{i1}^{b} P_R \right] v (\po )
\epsilon_{\mu} (p_W ) \\ \label{m2}
{\cal M}_{2,i=1,2} & = & g^2  \bar u (p_b ) \left[ l^{t}_{1i} P_R +
k^{t}_{1i} P_L \right]
\frac{ \not \! \pxi -\mxi }
     { \pxi^{2} -\mxi^{2}  - i \mxi \Gamma_{\tilde{\chi}^{\pm}_i}}
\left[ {\cal O}^{L}_{1i} P_L +{\cal O}_{1i}^{R} P_R  \right]
\gamma^{\mu} v (\po ) \epsilon_{\mu } (p_W ) \nonumber \\ \\ \label{m3}
{\cal M}_3 & = & - \frac{g^2}{\sqrt{2}} \bar u (p_b ) \gamma^{\mu }
\frac{1- \gamma_5 }{2}
\frac{ \not \! p_t +m_t }
     { p_{t}^{2} -m_{t}^{2}  - i m_t \Gamma_{t}}
\left[ b_{11}^{t} P_L + a_{11}^{t} P_R \right] v (\po )
\epsilon_{\mu } (p_W )
\end{eqnarray}
with $f_1(b) = \cos \theta_b$ and $f_2(b) = - \sin \theta_b$.
The decay width is given by:
\begin{eqnarray}
\Gamma(\tilde{t}_1 \to b W \tilde{\chi}^0_1) =
\frac{1}{2 m_{\tilde t_1}(2\pi )^5}
\int \frac{d^3 p_b}{2 E_b} \frac{d^3 p_W}{2 E_W} \frac{d^3
p_{\tilde{\chi}^0_1}}{2E_{\tilde \chi^0_1}}
\delta(p_{\tilde{t}_1} - p_b - p_W - p_{\tilde{\chi}^0_1})
|{\cal M}_1+{\cal M}_2+{\cal M}_3|^2
\end{eqnarray}
with
\begin{eqnarray}
{\cal M}_1 = {\cal M}_{1,1} + {\cal M}_{1,2}, \hspace{3mm}
{\cal M}_2 = {\cal M}_{2,1} + {\cal M}_{2,2}.
\end{eqnarray}

In order to complete the picture we will also rewrite the results of
\cite{hikasa} for the two body decay. They found, that the decay is
dominated by top--charm squark mixing, which is induced at one loop level.
In the limit $m_c \rightarrow 0$ only the left charm squark
contributes to this mixing.
The respective $\tilde t_1 - \tilde t_2 - \tilde c_L$ mixing
is in the basis of (\ref{mixing}) and in our notation given by
\begin{equation}
{\cal M}^{2}_{ \tilde t_1\tilde t_2  \tilde c_L }=\left(
\begin{array}{ccc}
m^{2}_{ \tilde t_1 } & 0 & \Delta_L \cos \theta_t + \Delta_R
\sin \theta_t \\
0 & m^{2}_{ \tilde t_2 } & - \Delta_L \sin \theta_t + \Delta_R \cos
\theta_t \\
\Delta_{L}^{\ast } \cos \theta_t + \Delta_{R}^{\ast} \sin \theta_t &
-\Delta_{L}^{\ast} \sin \theta_t + \Delta_{R}^{\ast} \cos \theta_t &
m^{2}_{ \tilde c_L}
\label{fcnmat}
\end{array} \right)
\end{equation}
The $\Delta_L$ ($\Delta_R$) are the mixing terms for the $\tilde t_L -
\tilde c_L $ ($\tilde t_R - \tilde c_L $) mixings with
\begin{eqnarray}
\Delta_L & = & - \frac{g^2}{16 \pi^2} \ln \left(\frac{M_X^2}{m_W^2} \right)
\frac{K^{\ast}_{tb} K_{cb} m_b^2 }{2 m_W^2 \cos^2 \beta }
( M_Q^2 + M_D^2 + M_{H_1}^{2} + |A_b|^2 )
\label{deltal} \\[0.2cm]
\Delta_R & = & \frac{g^2}{16 \pi^2} \ln \left(\frac{M_X^2}{m_W^2}
\right)
\frac{K^{\ast}_{tb} K_{cb} m_b^2 }{2 m_W^2 \cos^2 \beta } m_t A^{\ast}_{b}
\label{deltar}
\end{eqnarray}
where $M_X$ is a high scale which we assume to be the Planck mass to get a
maximal mixing. The $ M_Q ,\, M_D $ and
$ M_{H_1} $ are the squark-, down-squark and Higgs mass terms
and the $K_{tb} $ and $  K_{cb} $ are the respective elements of the
CKM matrix.

One gets \eq{deltal} and (\ref{deltar}) as one step solutions in
$\ln(M^2_P / M^2_W)$ of the renormalization group equation in the
framework of supergravity theories.  Note that one should stay away
from $A_b = 0$ because otherwise higer order terms in $\ln(M^2_P / M^2_W)$
would become important for $\Delta_R$. One should also note that in this
approximation $M_D$, $M_Q$ and $M_{H_1}$  can be evaluated at any scale
because the induced error would be of higher order. Therefore the expressions
should be treated as rough estimations giving the order of magnitude for
the mixing.

In the following
$\epsilon $ gives the size of the charm squark component of the lighter
stop, which we calculated numerically.
Therefore in this decay mode the charm-squark component
of the lighter stop couples with the charm quark and the LSP $\tilde
\chi_1^0 $ and the width is given by
\begin{equation}
\Gamma (\tilde t_1 \rightarrow c \tilde \chi_1^0 ) =
\frac{g^2}{16 \pi} \epsilon^2 |f^{c}_{11}|^2
m_{\tilde t_1 } \left( 1- \frac{m_{\tilde \chi_1^0}^{2}}{m_{\tilde t_1 }^{2}}
\right)^2
\end{equation}
where $f^c_{11} = \textstyle -\frac{2 \sqrt{2}}{3} \sin \theta_W N_{11}
               - \sqrt{2} ( \frac{1}{2} - \frac{2}{3} \sin^2 \theta_W)
                 \frac{N_{12}}{\cos \theta_W}$.

\section{Numerical Results}

In this Section we will first
describe that region of parameter space relevant for
our calculations. Then we will discuss typical decay widths of the three body
decay. In the last part we will compare these results for the three body
decay with those for the one loop
decay $\stchic$.

Since the decay $\stdrei$ becomes of interest if
$m_W + m_{\tilde \chi_{1}^{0}} + m_b < m_{\tilde t_1}
< m_{\tilde \chi_{1}^{+}} +m_b $ we show in \fig{region} different regions
in the $M_2 - \mu$~plane where this relation is valid. We show results for
two different values of $\tanb$ (2 and 30) and for stop masses of 170~GeV
and 220~GeV. The lower stop mass is relevant for an upgraded Tevatron
whereas the higher one is of interest for the LHC and
a 500~GeV $e^+ e^-$ Linear Collider. Clearly for the LHC higher stop masses
are also of interest. In such a case the region in the $M_2 - \mu$~plane
will be shifted to higher values of $M_2$ and the $M_2$ range would become
broader.

For fixing the parameters of the squark sector we have chosen the following
procedure: additional to $\tanb$ and $\mu$ we have used within the stop sector
$\mst$ and $\cost$ as input parameters.
For the sbottom sector we have fixed
$M_Q, M_D$ and $A_b$ as input parameters.
We have used this mixed set of parameters in
order to avoid unnatural parameters in the sbottom sector.  Note that
because of $SU(2)$ invariance $M_Q$ also appears in the stop mass matrix
(\eq{smatrix}). It can be seen by \eq{smatrix}, (\ref{offdiag}) and
(\ref{mixing}) that by varition of $\mu$ or
$\tanb$ for fixed $\mst$ and $\cost$ one also varies $A_t$ and $M_U$.
Therefore the mass of the heavier stop can be calculated
from this set of input parameters:
\begin{equation}
m^2_{\tilde{t}_2} =
     \frac{2 M^2_{Q} \textstyle
           +2 m^2_Z \cos 2 \beta (\frac{1}{2} - \frac{2}{3} \sin^2 \theta_W)
           +2 m^2_t
           - m^2_{\tilde{t}_1} (1 + \cos 2 \theta_t ) }
           { 1 - \cos 2 \theta_t }
\end{equation}
In the sbottom sector obviously the physical quantities $m_{\tilde{b}_i}$
and $\cos \theta_b$ changes with $\mu$ and $\tanb$.

\subsection{The Three Body Decay}

We shall now discuss the numerical results for the decay width
$\Gamma(\stdrei)$.  The results relevant for $\mst = 220$~GeV, relevant
for the Linear Collider and the LHC, are shown in \fig{gam220} and for
$\mst = 170$~GeV, relevant for an upgraded Tevatron, in \fig{gam170}.
For the other physical quantities we used $m_W=80$~GeV, $\sin^2 \theta_W =
0.23$, $m_b =5$~GeV and $ m_t=175 $~GeV.

First we shall focus on the case 220~GeV. As can be seen from \fig{region}
$M_2$ can vary between $\sim 210$~GeV and $\sim~270$ GeV. Within this small
region there is no significant change of the nature of the LSP and
the charginos for the allowed values of $\mu$. Therefore we have fixed
$M_2$ at 250 GeV. We have also found that our results depend only weakly
on the parameters of the sbottom sector. To be specific we have used
$M_Q = M_D = 500$~GeV and $A_b = -350$~GeV. We have checked that with these
choices of parameters the following relations hold:
$m_{\tilde{b}_1} + m_W > \mst$ and
$m_{\tilde{b}_2}, m_{\tilde{t}_2} < 1$~TeV.


In \figz{gam220}{a} we show the dependence of the decay width on $\tanb$
for $\cos \theta_t = 0.7$, $\mu = \pm 500$~GeV and $\mu = \pm 750$~GeV.
One can see that the decay width
varies between 0.18 and 1.65 KeV. For negative $\mu$ we have
a maximum at $\tan \beta \sim 20$ due to the positive interference
between the top (${\cal M}_3$) and chargino terms (${\cal M}_{2,i}$).
For small $\tan \beta$ and positive $\mu$ the behaviour is dominated
by the fact that the lighter chargino is nearly on mass shell. To
control this effect we have taken into account the decay widths in all
propagators.

In \figz{gam220}{b} we show the dependence of the decay width on $\cost$
for $\tan \beta = 20$, $\mu = \pm 500$~GeV and $\mu = \pm 750$.
As one can see the decay widths
varies between 10 eV and 0.78 KeV.
The maximum near $\cos \theta = 0.25$ is
due to the interference of the gaugino and higgsino parts in the squark
couplings. One can see that the decay width is slightly higher for positive
$\mu$ which results from different kinematics.


The now following discussion for the case $\mst = 170$~GeV will be
changed slightly. From \fig{region} one can see that the
region in $M_2 - \mu$~plane is smaller compared to the case above
and varies with $\tanb$.
Therefore we show only the dependence on $\cost$.
We show this for four different choices of $M_2, \mu$ and $\tanb$
(see Table 1) in \fig{gam170}. For the sbottom sector we have again taken
$M_Q=M_D=500$~GeV and $A_b=-350$~GeV.

The qualitative behaviour of this dependence is similar to that for
$m_{\tilde t_1}=220$~GeV. We also reach a maximum for the width for
$0< \cos \theta_t < 0.25$ by the same reason already mentioned for the
case $m_{\tilde t_1}=220$~GeV. But the width is in this case even smaller,
a few eV or even below.
This small width arises by the small difference of the masses
$ \Delta m = m_{\tilde t_1} - m_b - m_W - m_{\tilde \chi_1^0} $.
For our choice of parameters, $\Delta m$ varies between 0.6~GeV (scenario
a) and 2.4~GeV (scenario c).

\subsection{The Comparison of the Decay Modes}

We now will compare our results for the decay $\tilde t_1 \rightarrow
b W \tilde \chi_1^0$ with those of the decay $\tilde t_1 \rightarrow
c \tilde \chi_1^0$ The latter was calculated in \cite{hikasa}.
As already stated in Section 2 the used formula for the two body decay
gives a rough estimation for the order of magnitude. Therefore we
will mainly demonstrate the existence of parameter regions where one of
the decays clearly dominates.

Here we will follow the same procedure as in the last Section. Before
discussing our results in detail we will give some general remarks.
The crucial parameter for the width $\Gamma ( \tilde t_1 \rightarrow
c\tilde \chi_1^0 )$ is the size of the charm squark component $\epsilon$
of the physical stop.
We reached in some scenarios values for $\epsilon$ bigger than $0.1$.
$\epsilon$ will become big if ($i$) $m_{\tilde t_1 }$ and $m_{\tilde c_L}$
have almost the same size, $(ii)$ $\tan \beta $ becomes big ($\cos \beta$
small) which will enhance $\Delta_L$ and $\Delta_R$ $(iii)$
$\tan \theta_t \sim \Delta_L / \Delta_R$
which will maximize the ${\cal M}^2_{13}$ and ${\cal M}^2_{31}$
components of the mixing matrix
${\cal M}^2_{\tilde{t}_1 \tilde{t}_2 \tilde{c}_L}$ (\eq{fcnmat})
and $(iv)$ the parameters $M_D,\,M_Q,\,M_{H_1}$ and $A_b$ entering
$\Delta_L$ and $\Delta_R$ are big.

The charm squark mass is given by the
value of $M_Q$ and the contribution of the D-term and is with our choice of
parameters significantly higher than the stop mass
($m_{\tilde c_L} = 498.3$~GeV (497.1~GeV) for $\tan \beta =2 (30)$
 respectively). Therefore we do
not have an effect from the charm squark mass which could enhance
the width of the two body decay as mentioned above.

From \fig{region} one can see
that in the allowed regions $M_2$ is smaller than
$\mu$ in most of the cases. Therefore the lightest neutralino and
the lighter chargino will be mainly gaugino-like. Due to this fact
the influence of $\mu$ is mainly through phase space effects.
In the case that $\mu \leq M_2$ the coupling $\tilde \chi_1^0 \tilde c c $
will be small leading to an increase of the branching ratio of the
three body decay.

In \fig{br220} we consider the case $m_{\tilde t_1}= 220$~GeV.
As a typical example we take the case $\mu = -500$~GeV from
\fig{gam220} where we show the corresponding decay width of the
three body decay. \figz{br220}{a} shows the branching ratios as a
function of $\tan \beta$ for $\cos \theta_t = 0.7$. The reason
for the dominance of the two body decay for large $\tan \beta$ is
the above mentioned dependce of $\Delta_{L,R}$ on $1/ \cos \beta$.
Therefore the banching ratio for the three body decay is below
1\% for $\tan \beta > 30$. In \figz{br220}{b} we show the
branching ratio as function of $\cos \theta_t$ for $\tan \beta = 20$.
As already mentioned, $\epsilon$ will be maximized
if $\tan \theta_t \sim \Delta_L / \Delta_R $, which happens with our choice
of $A_b$ if $\cos \theta_t$ and $\sin \theta_t$ have the opposite sign and
$|\cos \theta_t |$ is big.  As we can see
this results in a strong dominance of the two body decay for $\cos \theta_t
< -0.4 (>0.7)$ whereas the three body decay is dominating near
$\cos \theta_t =0$.

The case $m_{\tilde t_1}=170$~GeV is shown in
\fig{br170} (the parameters are given in Table 1),
where we see a similar feature.
The decay $\tilde t_1 \rightarrow c \tilde \chi_1^0 $ becomes dominant
(or at least important in case of $\mu = -1000$ GeV and $\tan \beta = 2$)
if $|\cos \theta_t |$ becomes big. The dominance is stronger for $\tan \beta
=30 $ than for $\tan \beta =2$ as already explained. It is worthwhile to
mention that even for $\tan \beta = 30$ for very small $\cos \theta_t$
the three body decay may dominate resulting in this remarkable peak for the
respective branching fractions.

As main result we conclude, that there exists parameter regions where
either the three body decay or the two body decay dominates clearly.
This dominance may become so strong, that
the mentioned uncertainty is of no relevance in the respective parameter
region. Another important result is that we never got total decay widths
for the light stop bigger than 100~KeV. Therefore in all considered cases
the lifetime of the light stop will be larger than the hadronization scale.

\section{Conclusion}

We calculated the three body decay $\tilde t_1 \rightarrow b W \tilde
\chi_1^0 $ and compared these results with those for the two body decay
$\tilde t_1 \rightarrow c \tilde \chi_1^0 $ \cite{hikasa}.
These both decays will be the
competetive ones in that part of the parameter space accessible for either
the upgraded Tevatron, the LHC 
or a 500 GeV Linear Collider, where the two body
tree level decay $\tilde t_1 \rightarrow b \tilde \chi_{1}^{\pm} $ is
kinematically forbidden.

We found that the branching ratios are very sensitive to the choice
of the free parameters of the model.
Especially the stop mixing angle
$\theta_t $, the difference between the masses of the lighter stop
$m_{\tilde t_1}$ and the lefthanded charm squark $m_{\tilde c_L}$
and the value of $\tan \beta $ are crucial parameters, whereas
$M_2$ and $\mu$ are mainly important in order to specify the relevant regions
of the parameter space.
Depending on
the specific values of these parameters each decay mode may become the
dominant one and none of them should be neglected.

In case of a dominance of the three body decay, stop production
leads to the signature $2b + 2W + \not \! E$ which will
result in the final signature $(2-4)jets+(0-2)charged\,leptons+\not \! E $.
It is not trivial to answer the question, if a stop in this
parameter region is distinguishable from a top quark. Further investigations
especially of the differential widths
including Monte Carlo simulations are needed
to solve this problem.

Another important problem is that of hadronization of the produced stops.
We calculated a width in the range between 10~eV and 100~KeV.
One can clearly see that the lifetime of the light stop is bigger than
the hadronization time.
Therefore the mentioned Monte Carlo studies have also to address
all problems related to hadronization.

We have shown, that the three body decay $\tilde t_1 \rightarrow b W \tilde
\chi_1^0 $ is of major interest for stop physics at the upgraded Tevatron
as well as a 500 GeV Linear Collider and the LHC.
This decay cannot be neglected
in future investigations. But we also have shown that further investigations
are needed in order to give realistic predictions for experiments
at future colliders.

\section*{Acknowledgements}
We would like to thank X. Tata, A.~Bartl and W. Majerotto
for many helpful discussions.
T.W. also is gratefully to
the Departement of Physics and Astronomy of the University of Hawaii at Manoa,
the Departement of Physics and Astronomy of the University of Wisconsin
at Madison and the Institut f\"ur Theoretische Physik der Universit\"at
Wien, where part of this work was done,
for the kind hospitality and the pleasant atmosphere.
T.W. is supported by the Deutsche Forschungsgemeinschaft (DFG), and
W.P. by the ``Fonds zur F\"orderung der wissenschaftlichen Forschung''
of Austria, project no. P10843-PHY.

\section*{Appendix}
In this Appendix we give the full expressions of the squared amplitudes
$|\sum_n {\cal M}_n |^2$, with
${\cal M}_1:= {\cal M}_{1,1} +{\cal M}_{1,2}$ and
${\cal M}_2:= {\cal M}_{2,1} +{\cal M}_{2,2}$,
in terms of four-vector products $ p_l \cdot
p_k $ of the outer momenta of the bottom quark $p_b$, the $W$-boson $p_W$
and the lightest Neutralino $\po $. All sums run over $i,j=1,2$. The momenta
of the virtual particles are given by $\pbi = \pbj = p_b + \po$,
$\pxi = \pxj = \pw + \po $ and $ p_t = \pw + p_b$.
In this notation the squared amplitudes are given by
\begin{eqnarray}
|{\cal M}_{1}|^2 &=& 16 \sum_{i,j} \aiii\aiij \frac{1}{(\pbi^{2} -\mbi^{2}
+ i \mbi \Gamma_{\tilde b_i})}
\frac{1}{(\pbj^{2} -\mbj^{2}-i \mbj \Gamma_{\tilde b_j})} \nonumber \\
 & & \left\{ \left[ \frac{1}{m_{W}^{2}}\left( (p_b \cdot \pw )^2 + (\po \cdot
\pw )^2 +2 p_b \cdot \pw \po \cdot \pw \right) -\mb^{2} -\mo^{2} -2 p_b \cdot
\po \right] \times \right. \nonumber \\ & & \left.
\left[ (\aiki \aikj + \biki \bikj )p_b \cdot \po - ( \aiki \aikj -\biki
\bikj )\mb \mo \right] \right\} \\
|{\cal M}_{2}|^2 &=& \sum_{i,j} \frac{1}{(\pxi^{2} -\mxi^{2}
+ i \mxi \Gamma_{\tilde{\chi}^{\pm}_i})}
\frac{1}{(\pxj^{2} -\mxj^{2}-
i \mxj \Gamma_{\tilde{\chi}^{\pm}_j})} \nonumber \\ & &
4 ( \akii \akki \akij \akkj + \bkii \bkki \bkij \bkkj + \akii \akki \bkij \bkkj
+ \akii \akkj \bkij \bkki + \nonumber \\ & &
\akii \akij \bkki \bkkj + \bkii \bkki \akij \akkj
+ \bkii \bkkj \akij \akki + \bkii \bkij \akki \akkj ) \nonumber \\
 & & \Bigg[ p_b \cdot \po (\mo^{2} -\mw^{2} ) +4 \po \cdot \pw ( p_b \cdot \pw
+ p_b \cdot \po ) + \nonumber \\ & &
2 \mo^{2} (2 p_b \cdot \pw + \mo^{2} ) + \frac{2}{\mw^{2}}
\po \cdot \pw (2 p_b \cdot \po \pw \cdot \po - p_b \cdot \pw \mo^{2} )\Bigg]
\nonumber \\ & & +
12 ( \akii \akki \akij \akkj + \bkii \bkki \bkij \bkkj 
- \akii \akki \bkij \bkkj
+ \akii \akkj \bkij \bkki - \nonumber \\ & &
\akii \akij \bkki \bkkj - \bkii \bkki \akij \akkj
+ \bkii \bkkj \akij \akki - \bkii \bkij \akki \akkj ) \nonumber \\
 & & \mo \mb \mxi \mxj \nonumber \\ & & - 12 \left(\mxj
( \akii \akki \akij \akkj - \bkii \bkki \bkij \bkkj - \akii \akki \bkij \bkkj
+ \akii \akkj \bkij \bkki - \right. \nonumber \\ & & \left.
\akii \akij \bkki \bkkj + \bkii \bkki \akij \akkj
- \bkii \bkkj \akij \akki + \bkii \bkij \akki \akkj ) + i \leftrightarrow j
\right ) \nonumber \\ & & \mo (p_b \cdot \po +p_b \cdot \pw ) \nonumber \\ & &
+( \akii \akki \akij \akkj + \bkii \bkki \bkij \bkkj - \akii \akki \bkij \bkkj
- \akii \akkj \bkij \bkki + \nonumber \\ & &
\akii \akij \bkki \bkkj - \bkii \bkki \akij \akkj
- \bkii \bkkj \akij \akki + \bkii \bkij \akki \akkj ) \nonumber \\
 & & \mxi \mxj \left( \frac{8}{\mw^{2} } \po \cdot \pw p_b \cdot \pw + 4 \po
\cdot p_b \right ) \nonumber \\ & & +12
( \akii \akki \akij \akkj + \bkii \bkki \bkij \bkkj + \akii \akki \bkij \bkkj
- \akii \akkj \bkij \bkki - \nonumber \\ & &
\akii \akij \bkki \bkkj + \bkii \bkki \akij \akkj
- \bkii \bkkj \akij \akki - \bkii \bkij \akki \akkj ) \nonumber \\
& & \mo \mb ( \mw^{2} +\mo^{2} +2 \po \cdot \pw ) \nonumber \\ & & - \left(\mxj
( \akii \akki \akij \akkj - \bkii \bkki \bkij \bkkj - \akii \akki \bkij \bkkj
- \akii \akkj \bkij \bkki + \right. \nonumber \\ & & \left.
\akii \akij \bkki \bkkj + \bkii \bkki \akij \akkj
+ \bkii \bkkj \akij \akki - \bkii \bkij \akki \akkj ) + i \leftrightarrow j
\right ) \nonumber \\ & & \mb \left( 12 \po \cdot \pw + \frac{8}{\mw^{2}}
(\po \cdot \pw )^2 +4 \mo^{2} \right ) \\
|{\cal M}_3|^2 & = & \frac{1}{|\pst^{2} -\mt^{2}
- i m_t \Gamma_{t}|^2 } \left\{ 8|\alk |^{2}
| \ali +\bli |^{2}   \right. \nonumber \\ & &
\left[ \mb^{2} ( -6 \pw \cdot \po -3 p_b \cdot \po ) +\frac{4}{\mw^{2}}
\po \cdot \pw (p_b \cdot \pw )^2 - \right. \nonumber \\ & & \left.
\frac{2}{\mw^{2}} \mb^{2} \po \cdot p_b
\pw \cdot p_b +3 \mw^{2} \po \cdot p_b + 4 p_b \cdot \pw ( \po \cdot p_b
- \po \cdot \pw ) \right] \nonumber \\ & & +8 \mt^{2} |\alk |^{2}
| \ali -\bli |^{2}  \left( \frac{2}{\mw^{2}}
\po \cdot \pw p_b \cdot \pw + \po \cdot p_b \right) \nonumber \\ & &
\left. -16 \mt \mo (|\ali |^{2}  - |\bli |^{2} ) |\alk |^{2}  \left(
3 p_b \cdot \pw + \frac{2}{\mw^{2}} (p_b \cdot \pw )^2 - \mb^{2}
\right) \right\}\\
2\mbox{Re} ({\cal M}_1 {\cal M}_{2}^{\ast }) & = & 8 \mbox{Re} \sum_{i,j} \aiij
\frac{1}{(\pxi^{2} - \mxi^{2}
+ i \mxi \Gamma_{\tilde{\chi}^{\pm}_i})} \frac{1}{(\pbj^{2} - \mbj^{2}
-i \mbj \Gamma_{\tilde b_j})} \nonumber \\ & &
\left\{ (\akii \akki \aikj + \akii \bkki \bikj + \akki \bkii \bikj
+ \bkii \bkki \aikj ) \right. \nonumber \\ & &
\left[ \frac{\mo^{2} }{\mw^{2}} ( 2(p_b \cdot \pw )^2
+2 \po \cdot \pw p_b \cdot \pw ) - \frac{2}{\mw^{2}} \po \cdot p_b ( 2 (\po
\cdot \pw )^2 +\right. \nonumber \\ & &
2 \po \cdot \pw p_b \cdot \pw ) - 2 \mb^{2} (\mo^{2} + \pw \cdot
\po ) -2 p_b \cdot \po \pw \cdot \po +\nonumber \\ & & \left.
2 \mo^{2} (\po \cdot p_b + \pw \cdot p_b
) +2 \po \cdot p_b \pw \cdot p_b +4 (p_b \cdot \po )^2 \right] \nonumber \\ & &
- \mo \mxi (\akii \akki \aikj - \akii \bkki \bikj + \akki \bkii \bikj
- \bkii \bkki \aikj )  \nonumber \\ & & \left[ \frac{1}{\mw^{2}} (2(
p_b \cdot \pw )^2 +2 \po \cdot \pw p_b \cdot \pw ) -2\mb^{2} -2 p_b \cdot
\po \right] \nonumber \\ & &
+\left[ \mb \mo (\akii \akki \aikj - \akii \bkki \bikj - \akki \bkii \bikj
+ \bkii \bkki \aikj ) \right. \nonumber \\ & &
+\left. \mb \mxi (\akii \akki \aikj + \akii \bkki \bikj - \akki \bkii \bikj
- \bkii \bkki \aikj ) \right] \nonumber \\ & &
\left. \left[ \frac{1}{\mw^{2}} (2(
\po \cdot \pw )^2 +2 \po \cdot \pw p_b \cdot \pw ) -2\mo^{2} -2 p_b \cdot
\po \right] \right\} \\
2\mbox{Re} ({\cal M}_{1}^{\ast} {\cal M}_3) & = & 8 \mbox{Re} \sum_i \alk \aiii
\frac{1}{(\pst^{2} - \mt^{2} - i m_t \Gamma_{t})} \frac{1}{(\pbi^{2} - \mbi^{2}
+i \mbi \Gamma_{\tilde b_i})} \nonumber \\ & &
\left\{ (\ali \aiki + \bli \biki + \aiki \bli
+ \ali \biki ) \right. \nonumber \\ & &
\left[ -\frac{\mb^{2} }{\mw^{2}} ( 2(\po \cdot \pw )^2
+2 \po \cdot \pw p_b \cdot \pw ) + \frac{2}{\mw^{2}} \po \cdot p_b ( 2 (p_b
\cdot \pw )^2 +\right. \nonumber \\ & &
2 \po \cdot \pw p_b \cdot \pw ) + 2 \mo^{2} (\mb^{2} + \pw \cdot
p_b ) +2 p_b \cdot \po \pw \cdot p_b - \nonumber \\ & & \left.
2 \mb^{2} (\po \cdot \pw + \po \cdot p_b
) -2 \po \cdot p_b \pw \cdot \po -4 (p_b \cdot \po )^2 \right] \nonumber \\ & &
+ \mb \mt (\ali \aiki  + \bli \biki - \aiki \bli
- \ali \biki )  \nonumber \\ & & \left[ \frac{1}{\mw^{2}} (2(
\po \cdot \pw )^2 +2 \po \cdot \pw p_b \cdot \pw ) -2\mo^{2} -2 p_b \cdot
\po \right] \nonumber \\ & &
-\left[ \mt \mo (\ali \aiki  -  \bli \biki - \aiki \bli
+ \ali \biki ) \right. \nonumber \\ & &
+\left. \mb \mo (\ali \aiki  -  \bli \biki + \aiki \bli
- \ali \biki ) \right] \nonumber \\ & &
\left. \left[ \frac{1}{\mw^{2}} (2(
p_b \cdot \pw )^2 +2 \po \cdot \pw p_b \cdot \pw ) -2\mb^{2} -2 p_b \cdot
\po \right] \right\} \\
2\mbox{Re} ({\cal M}_{2}^{\ast} {\cal M}_{3}) & = & 8 \mbox{Re}\sum_{i} \alk
\frac{1}{(\pst^{2} - \mt^{2} - i m_t \Gamma_{t})} \frac{1}{(\pxi^{2} - \mxi^{2}
+ i \mxi \Gamma_{\tilde{\chi}^{\pm}_i})} \nonumber \\ & &
\left\{
(\akii \akki \ali + \bkii \akki \ali + \akii \bkki \bli + \bkii \bkki \bli
\right. \nonumber \\ & &
+ \akii \bkki \ali + \bkii \bkki \ali + \akii \akki \bli + \bkii \akki \bli )
\nonumber \\ & &
\left[ 2 \frac{\mb^{2} }{\mw^{2}} (\po \cdot \pw )^2 -\mo^{2} \mb^{2} -
\po \cdot p_b \mw^{2} -\right. \nonumber \\ & &
\frac{2}{\mw^{2}} p_b \cdot \pw ( 2 \po \cdot p_b
\po \cdot \pw - p_b \cdot \pw \mo^{2} )  \nonumber \\ & &
+ 2 p_b \cdot \po \pw \cdot \po + p_b \cdot \pw \mo^{2} +2 p_b \cdot \pw
p_b \cdot \po + \nonumber \\ & & \left.
\po \cdot \pw \mb^{2} +4 (\po \cdot p_b )^2 +4 p_b \cdot \pw
\po \cdot \pw \right] \nonumber \\ & & +\mxi \mt
(\akii \akki \ali + \bkii \akki \ali + \akii \bkki \bli + \bkii \bkki \bli
\nonumber \\ & &
- \akii \bkki \ali - \bkii \bkki \ali - \akii \akki \bli - \bkii \akki \bli )
\nonumber \\ & & \left( \frac{2}{\mw^{2}} p_b \cdot \pw \po \cdot \pw
+p_b \cdot \po
\right) \nonumber \\ & & -3\mo \mt
(\akii \akki \ali + \bkii \akki \ali - \akii \bkki \bli - \bkii \bkki \bli
\nonumber \\ & &
+ \akii \bkki \ali + \bkii \bkki \ali - \akii \akki \bli - \bkii \akki \bli )
\nonumber \\ & & (p_b \cdot \po + p_b \cdot \pw ) \nonumber \\ & & - \mo \mxi
(\akii \akki \ali + \bkii \akki \ali - \akii \bkki \bli - \bkii \bkki \bli
\nonumber \\ & &
- \akii \bkki \ali - \bkii \bkki \ali + \akii \akki \bli + \bkii \akki \bli )
\nonumber \\ & & \left( \frac{2}{\mw^{2}} (p_b \cdot \pw )^2 +\mb^{2} +3
p_b \cdot \pw \right) \nonumber \\ & & - \mb \mt
(\akii \akki \ali - \bkii \akki \ali + \akii \bkki \bli - \bkii \bkki \bli
\nonumber \\ & &
- \akii \bkki \ali + \bkii \bkki \ali - \akii \akki \bli + \bkii \akki \bli )
\nonumber \\ & & \left( \frac{2}{\mw^{2}} (\po \cdot \pw )^2 +\mo^{2}
+3 \po \cdot \pw \right) \nonumber \\ & & - 3 \mxi \mb
(\akii \akki \ali - \bkii \akki \ali + \akii \bkki \bli - \bkii \bkki \bli
\nonumber \\ & &
+ \akii \bkki \ali - \bkii \bkki \ali + \akii \akki \bli - \bkii \akki \bli )
\nonumber \\ & & ( \po \cdot p_b + \po \cdot \pw ) \nonumber \\ & & + \mb \mo
(\akii \akki \ali - \bkii \akki \ali - \akii \bkki \bli + \bkii \bkki \bli
\nonumber \\ & &
- \akii \bkki \ali + \bkii \bkki \ali + \akii \akki \bli - \bkii \akki \bli )
\nonumber \\ & & \left( 3 \mw^{2} + 3 \po \cdot \pw +3 p_b \cdot \pw +\po \cdot
p_b + \frac{2}{\mw^{2}} \po \cdot \pw p_b \cdot \pw \right ) \nonumber \\ & &
+ 3 \mxi \mb \mo \mt
(\akii \akki \ali - \bkii \akki \ali - \akii \bkki \bli + \bkii \bkki \bli
\nonumber \\ & & \left.
+ \akii \bkki \ali - \bkii \bkki \ali - \akii \akki \bli + \bkii \akki \bli )
\right\}
\end{eqnarray}
Here the $a_{nm}^{k}$ and $b_{nm}^{k}$ are coupling constants and given by
\cite{tata1}:
\[ \begin{array}{ r c l r c l }
a_{11}^{1} & = & -\frac{g}{\sqrt{2}} \cos \theta_b \cos \theta_t & & & \\
a_{11}^{2} & = & \frac{g}{\sqrt{2}} \sin \theta_b \cos \theta_t & & & \\
a_{12}^{i} & = & \frac{g}{2} \left( a_{i1}^{b} + b_{i1}^{b} \right),&
b_{12}^{i} & = & \frac{g}{2} \left( a_{i1}^{b} - b_{i1}^{b} \right) \\
a_{21}^{i } & = & \frac{g}{2} \left( l^{t}_{1i} + k^{t}_{1i} \right),&
b_{21}^{i } & = & \frac{g}{2} \left( l^{t}_{1i} - k^{t}_{1i} \right) \\
a_{22}^{i} & = & -\frac{g}{2} \left( {\cal O}^{L}_{1i} +
{\cal O}^{R}_{1i} \right), &
b_{22}^{i} & = & -\frac{g}{2} \left( {\cal O}^{L}_{1i} -
{\cal O}^{R}_{1i} \right)  \\
a_{31} & = & \frac{g}{2} \left( a_{11}^{t} + b_{11}^{t} \right),&
b_{31} & = & \frac{g}{2} \left( a_{11}^{t} - b_{11}^{t} \right) \\
a_{32} & = & - \frac{g}{2 \sqrt{2}} & & &
\end{array} \]

\newpage
\section*{Figure Captions}
{\bf Fig. 1:} \\
\refstepcounter{figure}
\label{graph}
Feynman graphs for the decay $\tilde t_1 \rightarrow b W \tilde \chi_1^0 $
related to the matrix elements ${\cal M}_{1,i=1,2},\, {\cal M}_{2,i=1,2}$
and ${\cal M}_3$ \eq{m1}--\eq{m3}.
The arrow of the fermionic lines defines a fermion flow
and is not necessarily identical with the momentum flow used in our
calculations.\\[0.4cm]
{\bf Fig. 2:} \\
\refstepcounter{figure}
\label{region}
Regions in the $\mu - M_2$ plane for $ m_{\tilde \chi_1^+} +m_b >
m_{\tilde t_1} > m_{\tilde \chi_1^0 } +m_b +m_W $ for
$m_W = 80$~GeV, $m_b = 5$~GeV, (I) $m_{\tilde t_1}=
220$ GeV and (II) $m_{\tilde t_1}=170$ GeV. Fig.a) is for the case $\tan
\beta =2$, whereas b) shows the case $\tan \beta =30$. The shaded region
will be probed by LEP2 assuming that signals from charginos with a mass
smaller than 90~GeV will be observable there.\\[0.4cm]
{\bf Fig. 3:} \\
\refstepcounter{figure}
\label{gam220}
Total width for the decay $\tilde t_1 \rightarrow bW \tilde \chi_1^0$ for
$m_{\tilde t_1}=220$ GeV. Fig.a) shows the width as a function
of $\tan \beta$ for
$\cos \theta_t=0.7$ and Fig.b) that of $\cos \theta_t$
for $\tan \beta = 20$. We have taken $M_2 = 250$~GeV and different values of
$\mu$: $\mu = -750$~GeV (solid line), $\mu = 750$~GeV (short dashed line),
$\mu = -500$~GeV (long dashed line) and $\mu = 500$~GeV (dotted dashed line).
The other parameters are given in the text. \\[0.4cm]
{\bf Fig. 4:} \\
\refstepcounter{figure}
\label{gam170}
Total width for the decay $\tilde t_1 \rightarrow bW \tilde \chi_1^0$
as a function of $\cos \theta_t$ for $m_{\tilde t_1}=170$~GeV. The parameters
(see also tab.1) are for a) (solid line)
$\mu=-500$~GeV, $M_2=165$~GeV and $\tan \beta=2 $,
for b) (short dashed line)
$\mu=500$~GeV, $M_2=165$~GeV and $\tan \beta=2 $,
for c) (long dashed line)
$\mu=-1000$~GeV, $M_2=166$~GeV and $\tan \beta=30 $
and for d) (dotted-dashed line)
$\mu=1000$~GeV, $M_2=167$~GeV and $\tan \beta=30 $.
\\[0.4cm]
{\bf Fig. 5:} \\
\refstepcounter{figure}
\label{br220}
The branching ratios for the decays $\tilde t_1 \rightarrow b W \tilde
\chi_1^0 $ (solid line) and $\tilde t_1 \rightarrow c \tilde \chi_1^0$
(dashed line) for $m_{\tilde t_1}=220$ GeV, $M_2 = 250$~GeV and
$\mu = -500$~GeV. The other parameters are explained in the text.
Fig.~a shows the dependence on $\tan \beta$ whereas Fig.~b shows the
dependence on $\cos \theta_t$. \\[0.4cm]
{\bf Fig. 6:} \\
\refstepcounter{figure}
\label{br170}
The branching ratios for the decays $\tilde t_1 \rightarrow b W \tilde
\chi_1^0 $ (solid line) and $\tilde t_1 \rightarrow c \tilde \chi_1^0$
(dashed line) for $m_{\tilde t_1}=170$ GeV and all other parameters
as in \fig{gam170} (see also tab.1). In case of scenario c) we also show the
branching ratio for the (in this case accessible) decay $\tilde t_1
\rightarrow c \tilde \chi_2^0$ (dotted-dashed line). Notice that we dropped
scenario d) due to the fact that there are no visible differences
between the scenarios c) and d).
\newpage
\section*{Tabular 1}
\begin{tabular}{ c c c c c c c }
\hline  \hline \\
scenario & $\mu $ & $ M_2$ & $\tan \beta $ & $m_{\tilde \chi_1^0} $ &
$m_{\tilde \chi_2^0} $ & $m_{\tilde \chi_1^+} $ \\  \hline \\
a) & -500 & 165 & 2 & 84.4 & 171.8 & 171.7 \\
b) &  500 & 165 & 2 & 83.4 & 169.2 & 169.2 \\
c) & -1000 & 166 & 30 & 82.6 & 165.4 & 165.4 \\
d) & 1000 & 167 & 30 & 82.8 & 165.5 &  165.5 \\ \hline \hline
\end{tabular}
\newline
\vspace{1.0cm} \\
{\bf Tabular caption:}\\
Parameters used in the scenarios for the case $m_{\tilde t_1}=170$ GeV.
Additional we show the respective values for the masses (in GeV) of the two
lighter neutralinos and the lighter chargino. The other parameters are
fixed at $M_Q=M_D=500$~GeV and $A_b=-350$~GeV.

\begin{figure}
\vglue +1.5 true cm
 \unitlength 1mm
 \begin{picture}(175,240)
   \put(38,228){\large $\tilde{t}_1$}
   \put(106,215){\large $\tilde{\chi}^0_1$}
   \put(56,236){\large $W^+$}
   \put(75,228){\large $\tilde{b}_{1,2}$}
   \put(106,235){\large $b$}
   \put(38,161){\large $\tilde{t}_1$}
   \put(107,168){\large $W^+$}
   \put(107,148){\large $\tilde{\chi}^0_1$}
   \put(75,162){\large $\tilde{\chi}^-_{1,2}$}
   \put(60,168){\large $b$}
   \put(38,94){\large $\tilde{t}_1$}
   \put(55,100){\large $\tilde{\chi}^0_1$}
   \put(107,100){\large $W^+$}
   \put(75,94){\large $t$}
   \put(107,80){\large $b$}
   \put(57,50){\Large \bf Fig. 1 }
\put(20,70){\psfig{file=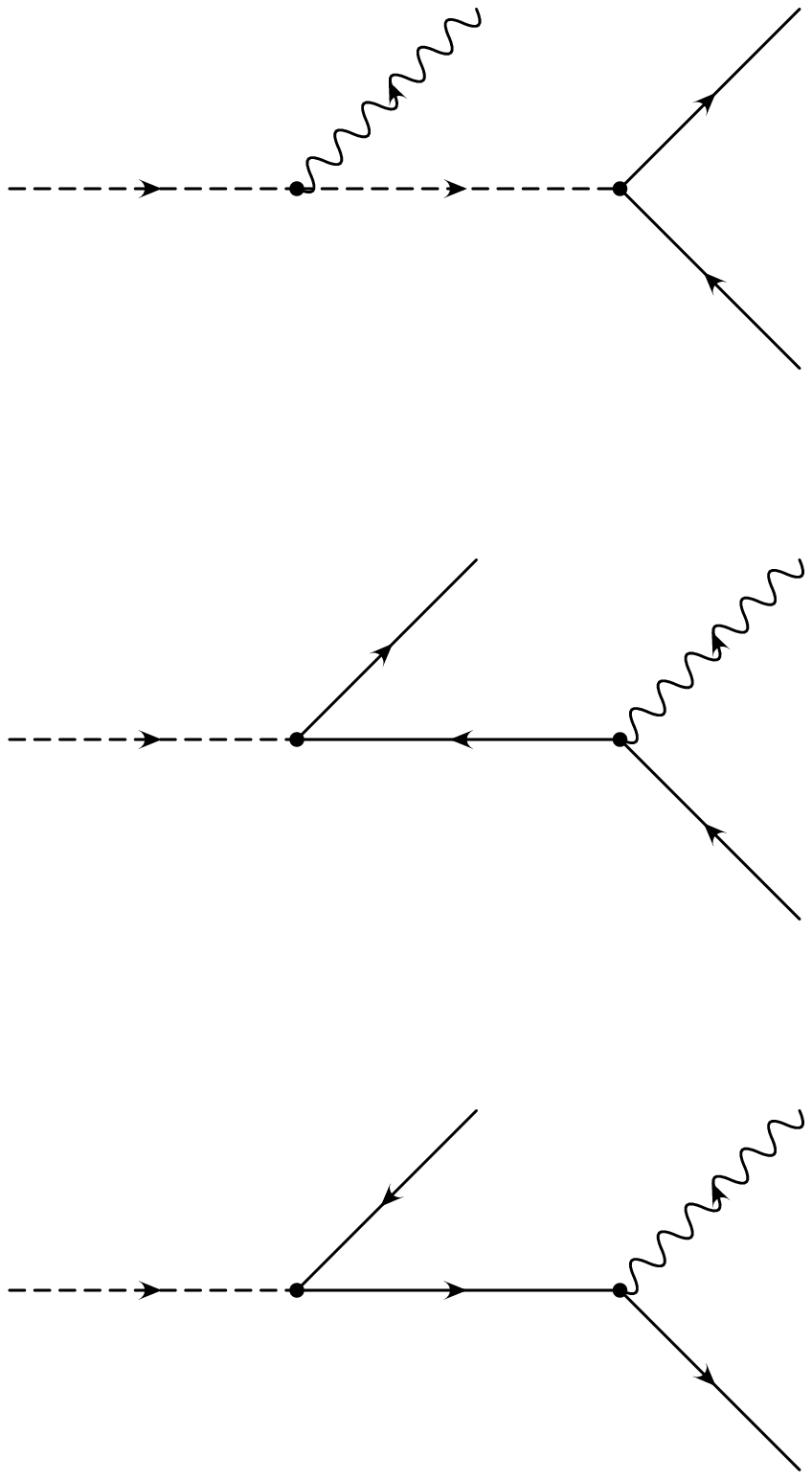,height=18cm}}
 \end{picture}
\end{figure}

\begin{figure}
\vglue +1.5 true cm
 \unitlength 1mm
 \begin{picture}(175,240)
  \put(0,249){\large \bf a)}
   \put(5,236){\large $M_2$}
   \put(1.5,231){(GeV)}
   \put(80,153.5){{\large $\mu$} (GeV)}
  \put(0,147){\large \bf b)}
   \put(5,136){\large  $M_2$}
   \put(1.5,131){(GeV)}
   \put(80,53){{\large $\mu$} (GeV)}
\put(4,42){\psfig{file=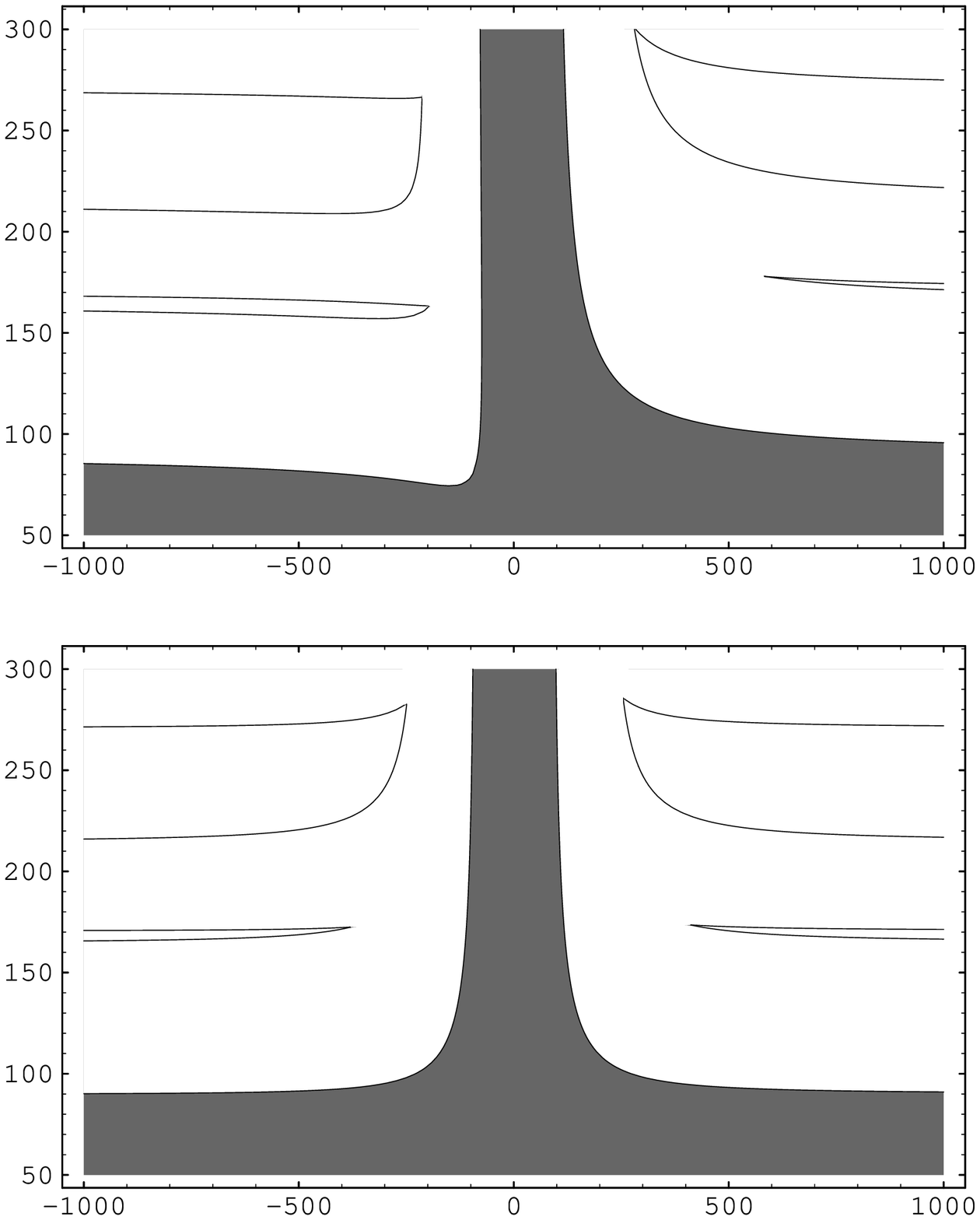,height=22cm}}
   \put(77,35){\Large \bf Fig. 2 }
   \put(45,223){\large I }
   \put(135,226){\large I }
   \put(45,202){\large II }
   \put(44,204){\vector(-1,-2){2.2} }
   \put(135,204){\large II }
   \put(139.5,207){\vector(1,-2){2} }
   \put(45,123){\large I }
   \put(135,123){\large I }
   \put(45,103.5){\large II }
   \put(44,105.5){\vector(-1,-2){2.1} }
   \put(135,103){\large II }
   \put(139.5,106.1){\vector(1,-2){2.1} }
 \end{picture}
\end{figure}

\begin{figure}
\vglue -2 true cm
 \unitlength 1mm
 \begin{picture}(175,240)
   \put(19,215){\large $\Gamma(\tilde{t}_1 \to b W \tilde{\chi}^0_1)$}
   \put(19,210){(keV)}
   \put(82,53){\large $\tan \beta$ }
   \put(66,20){\Large \bf Fig. 3a }
\put(4,32){\psfig{file=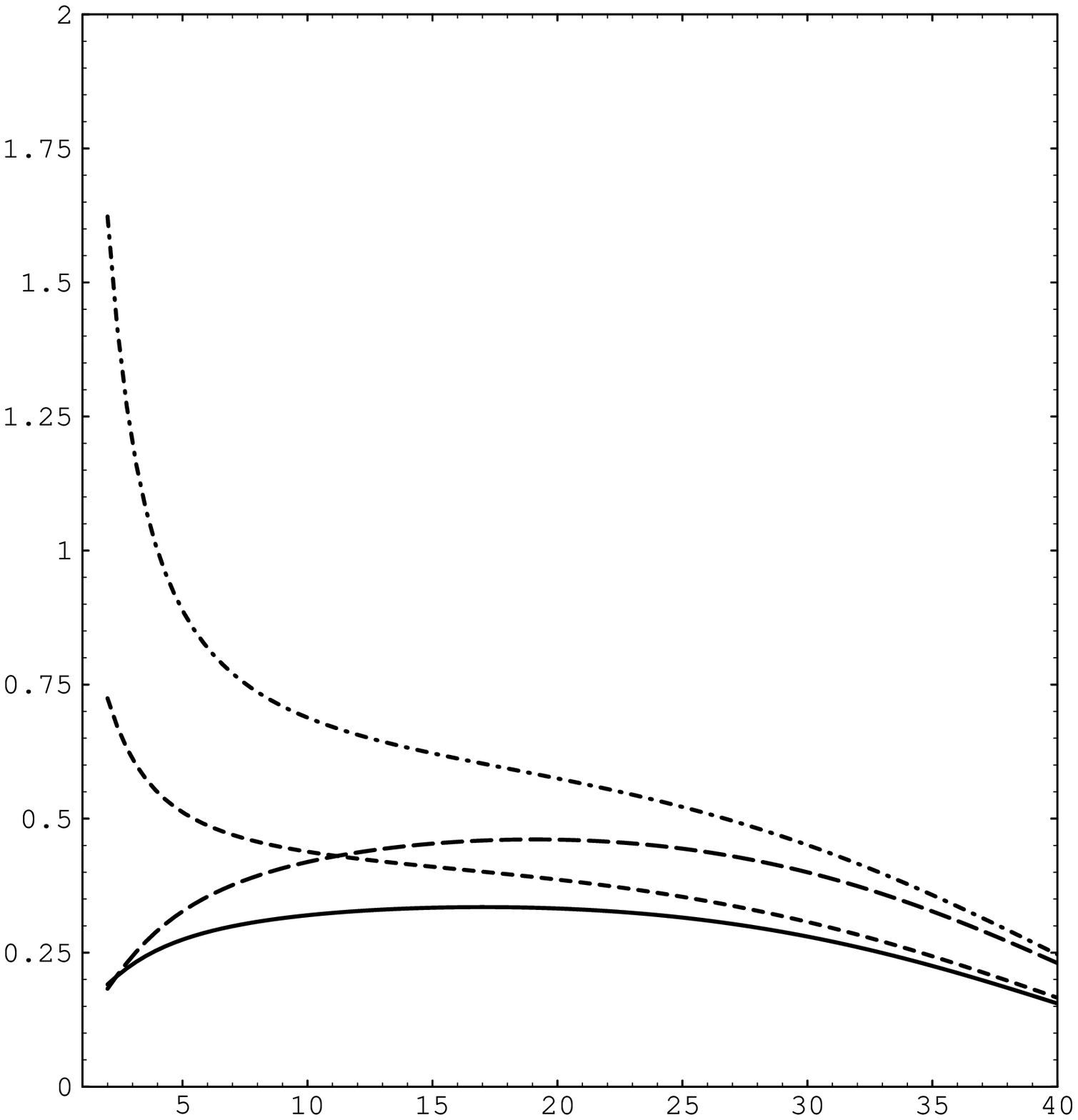,height=22cm}}
 \end{picture}
\end{figure}

\begin{figure}
\vglue -2 true cm
 \unitlength 1mm
 \begin{picture}(175,240)
   \put(19,215){\large $\Gamma(\tilde{t}_1 \to b W \tilde{\chi}^0_1)$}
   \put(19,210){(keV)}
   \put(84,53){\large $\cos \theta_t$ }
   \put(66,20){\Large \bf Fig. 3b }
\put(4,32){\psfig{file=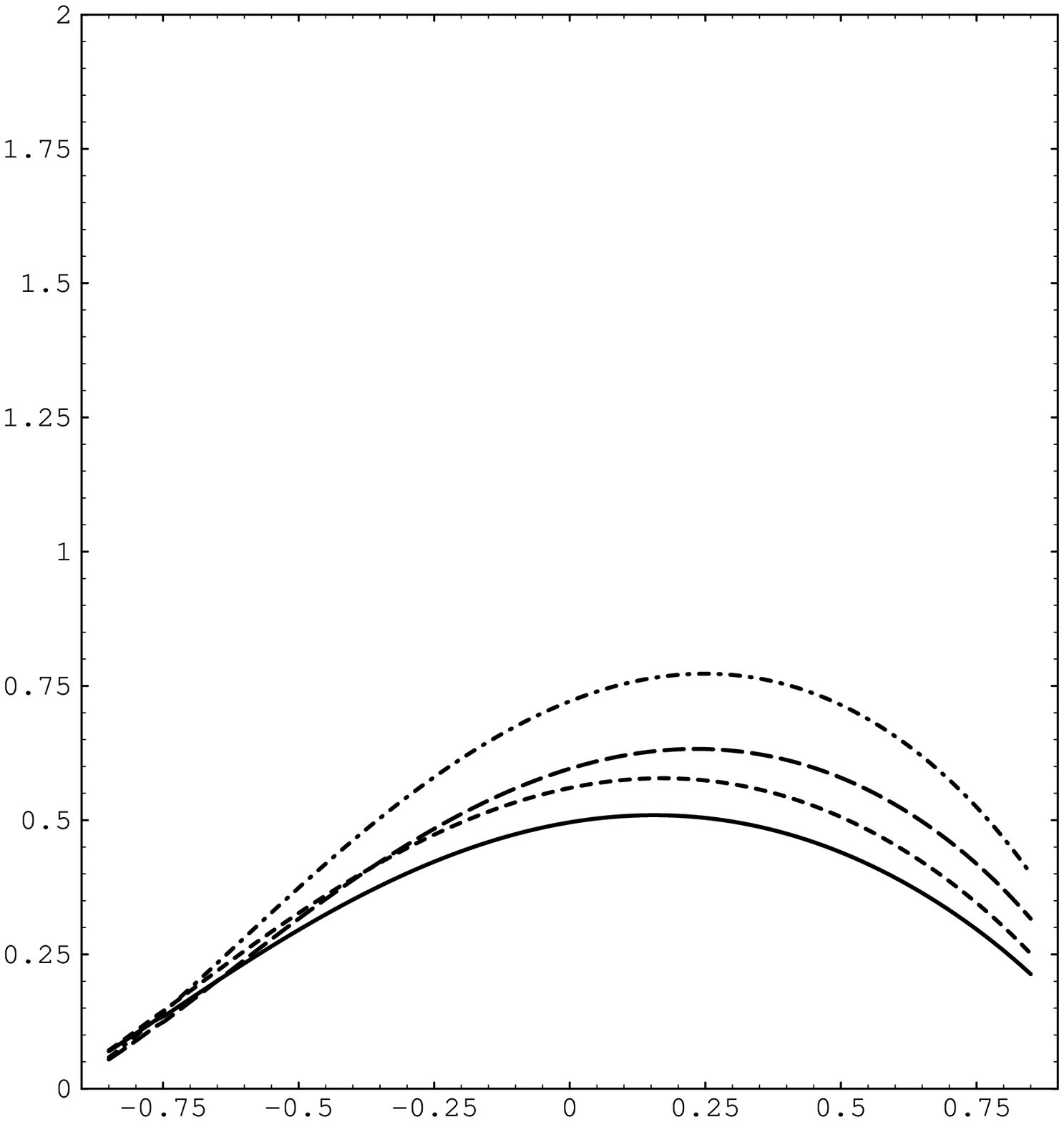,height=22cm}}
 \end{picture}
\end{figure}

\begin{figure}
\vglue -2 true cm
 \unitlength 1mm
 \begin{picture}(175,240)
   \put(25,215){\large $\Gamma(\tilde{t}_1 \to b W \tilde{\chi}^0_1)$}
   \put(25,210){(GeV)}
   \put(84,50){\large $\cos \theta_t$ }
   \put(66,20){\Large \bf Fig. 4 }
\put(4,32){\psfig{file=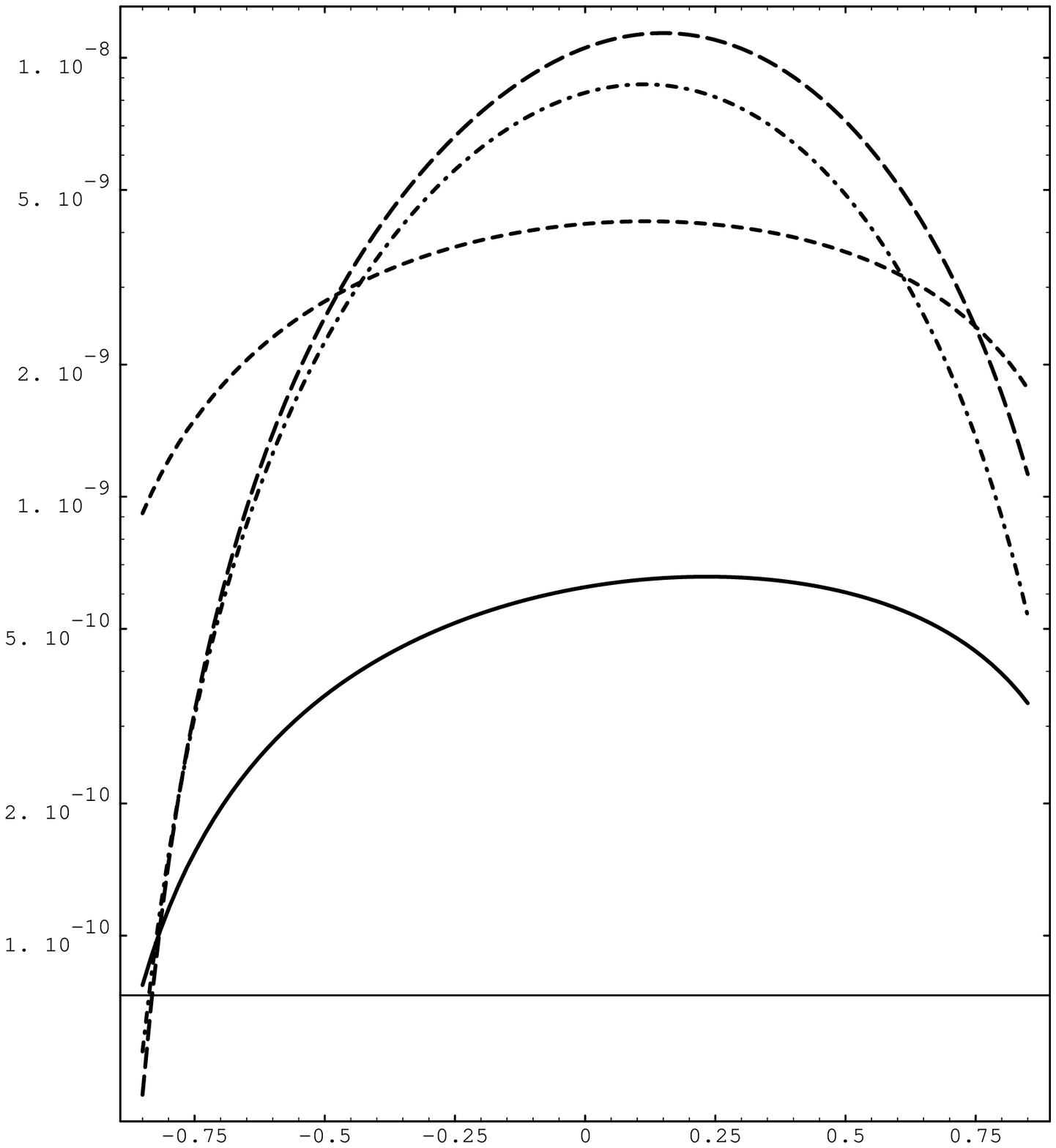,height=21.5cm}}
 \end{picture}
\end{figure}

\begin{figure}
\vglue -2 true cm
 \unitlength 1mm
 \begin{picture}(175,240)
   \put(35,215){\large $BR(\tilde{t}_1 \to b W \tilde{\chi}^0_1)$}
   \put(35,70){\large $BR(\tilde{t}_1 \to c \tilde{\chi}^0_1)$}
   \put(82,52){\large $\tan \beta$ }
   \put(66,20){\Large \bf Fig. 5a }
\put(4,32){\psfig{file=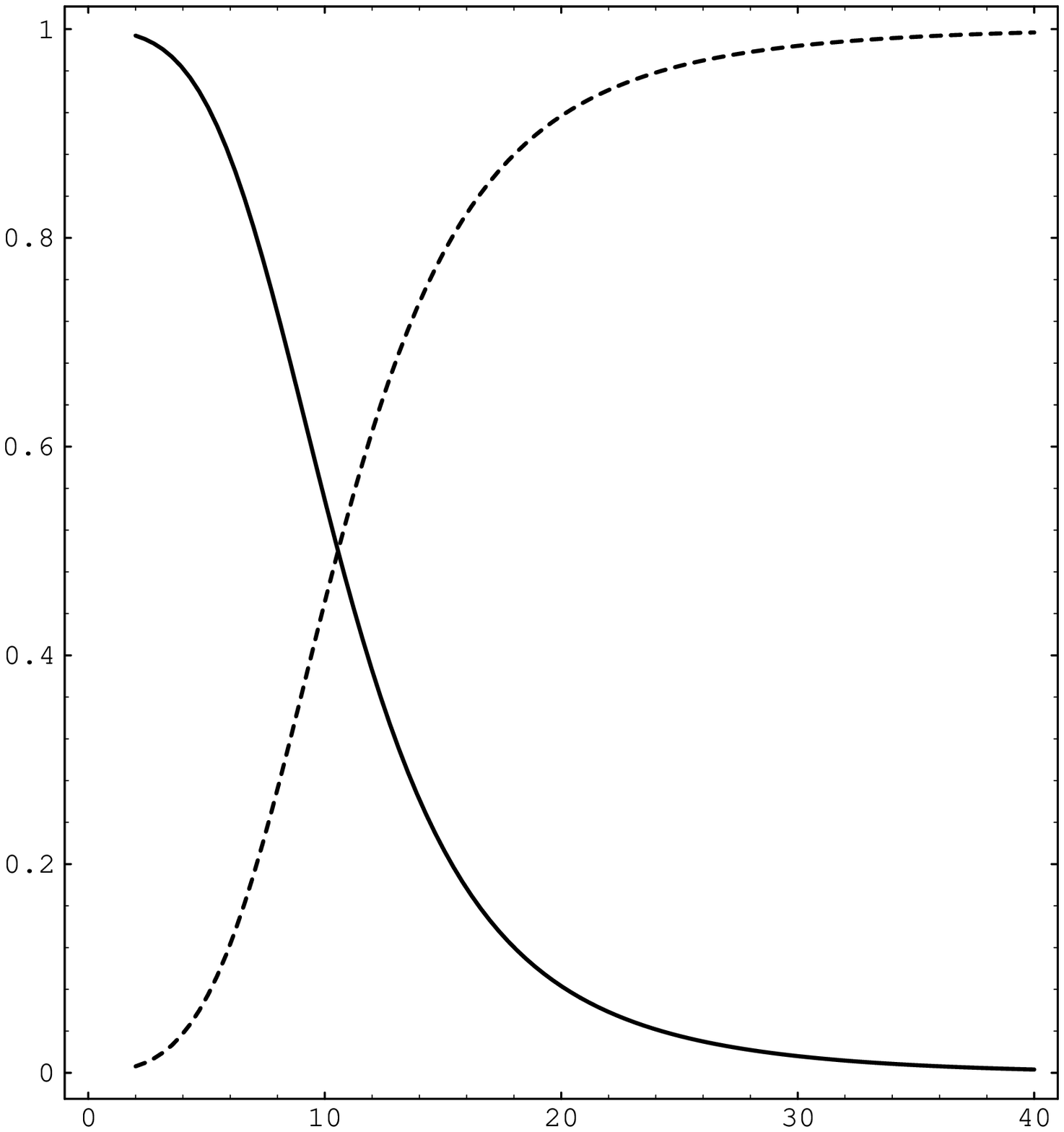,height=22cm}}
 \end{picture}
\end{figure}

\begin{figure}
\vglue -2 true cm
 \unitlength 1mm
 \begin{picture}(175,240)
   \put(33,70){\large $BR(\tilde{t}_1 \to b W \tilde{\chi}^0_1)$}
   \put(33,215){\large $BR(\tilde{t}_1 \to c \tilde{\chi}^0_1)$}
   \put(82,52){\large $\cos \theta_t$ }
   \put(66,20){\Large \bf Fig. 5b }
\put(4,32){\psfig{file=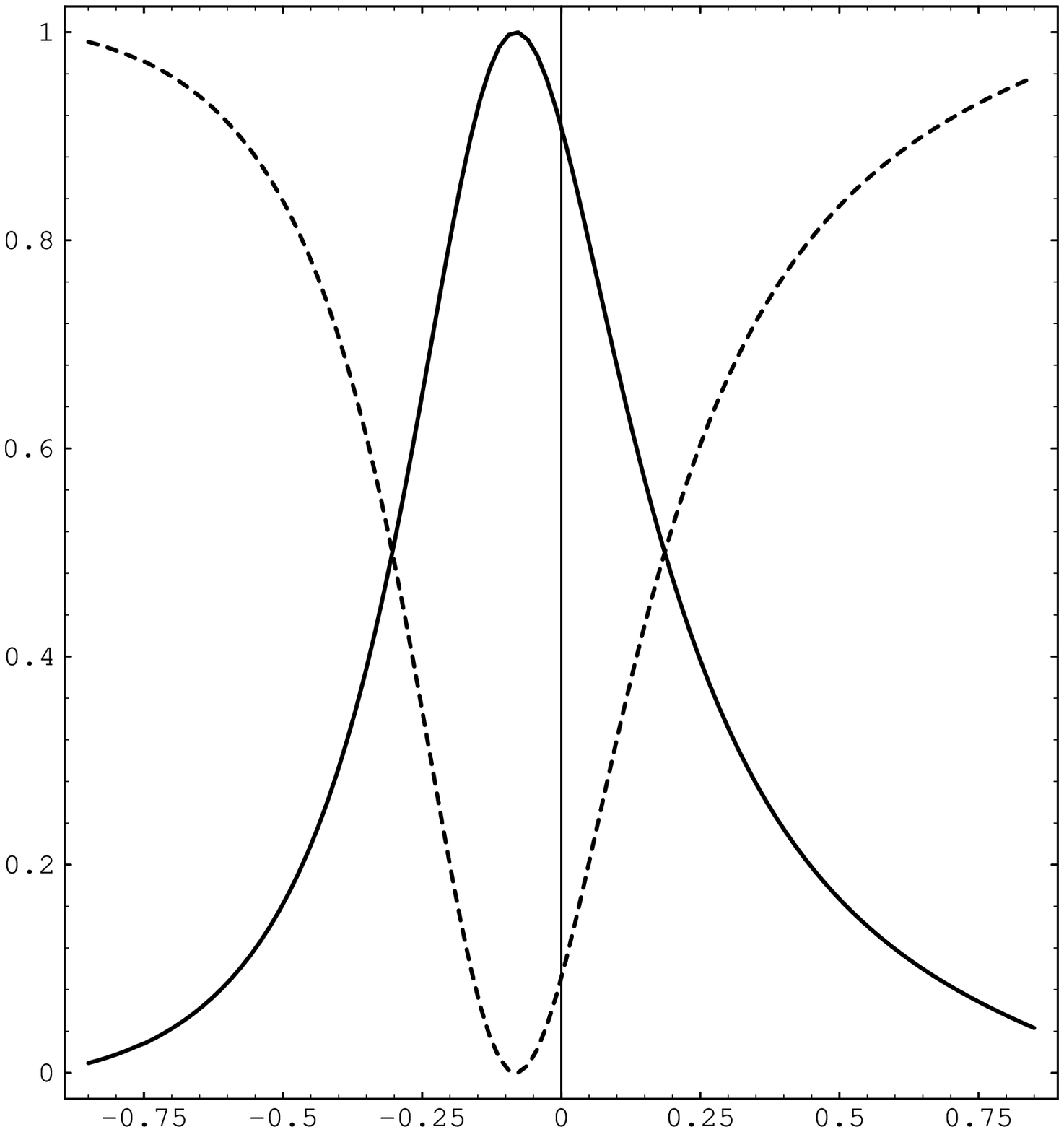,height=22cm}}
 \end{picture}
\end{figure}

\begin{figure}
\vglue -2 true cm
 \unitlength 1mm
 \begin{picture}(175,240)
   \put(17,70){\large $BR(\tilde{t}_1 \to b W \tilde{\chi}^0_1)$}
   \put(17,215){\large $BR(\tilde{t}_1 \to c \tilde{\chi}^0_1)$}
   \put(82,52){\large $\cos \theta_t$ }
   \put(66,20){\Large \bf Fig. 6a }
\put(4,32){\psfig{file=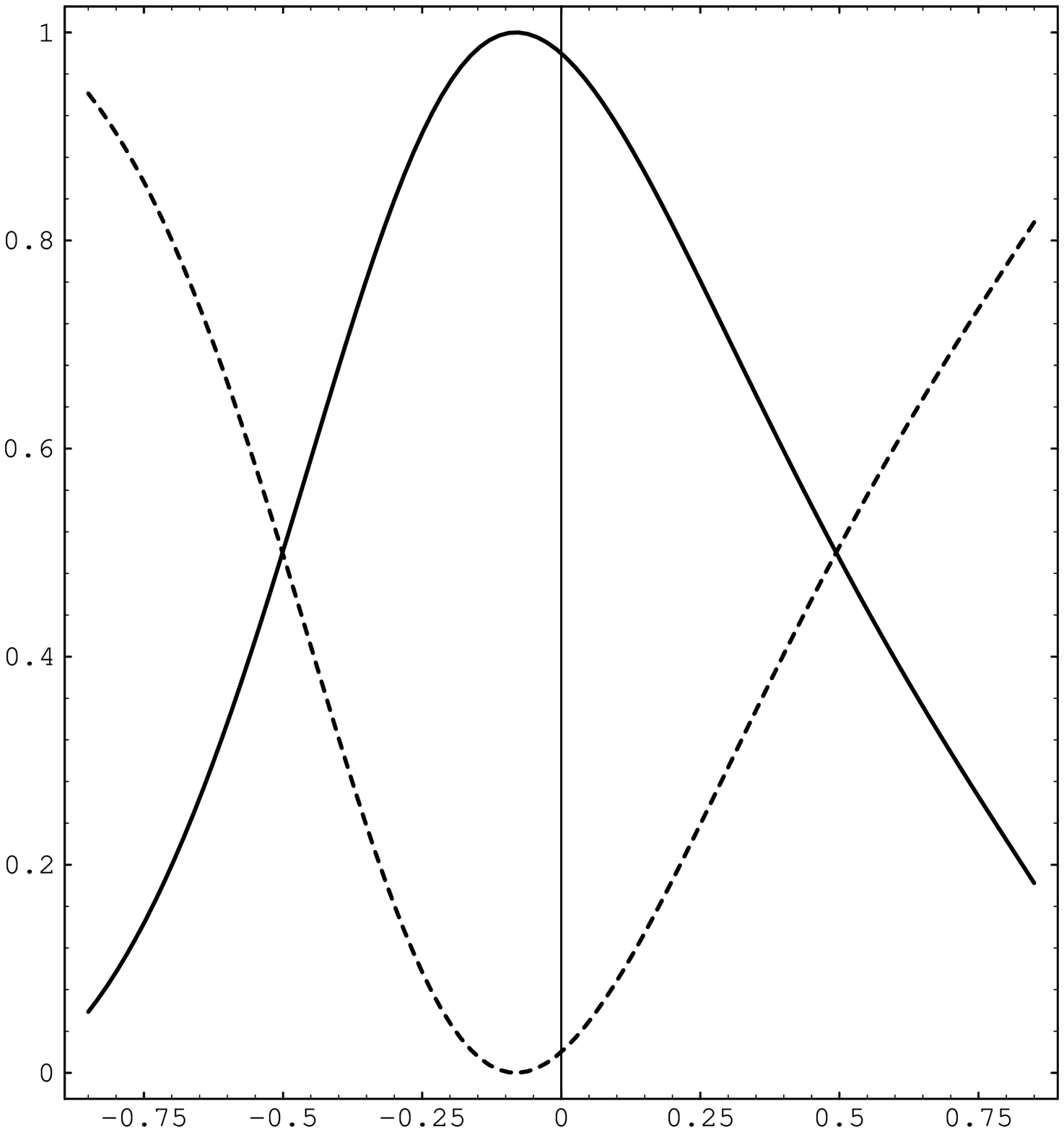,height=22cm}}
 \end{picture}
\end{figure}

\begin{figure}
\vglue -2 true cm
 \unitlength 1mm
 \begin{picture}(175,240)
   \put(16,215){\large $BR(\tilde{t}_1 \to b W \tilde{\chi}^0_1)$}
   \put(18,72){\large $BR(\tilde{t}_1 \to c \tilde{\chi}^0_1)$}
   \put(82,52){\large $\cos \theta_t$ }
   \put(66,20){\Large \bf Fig. 6b }
\put(4,32){\psfig{file=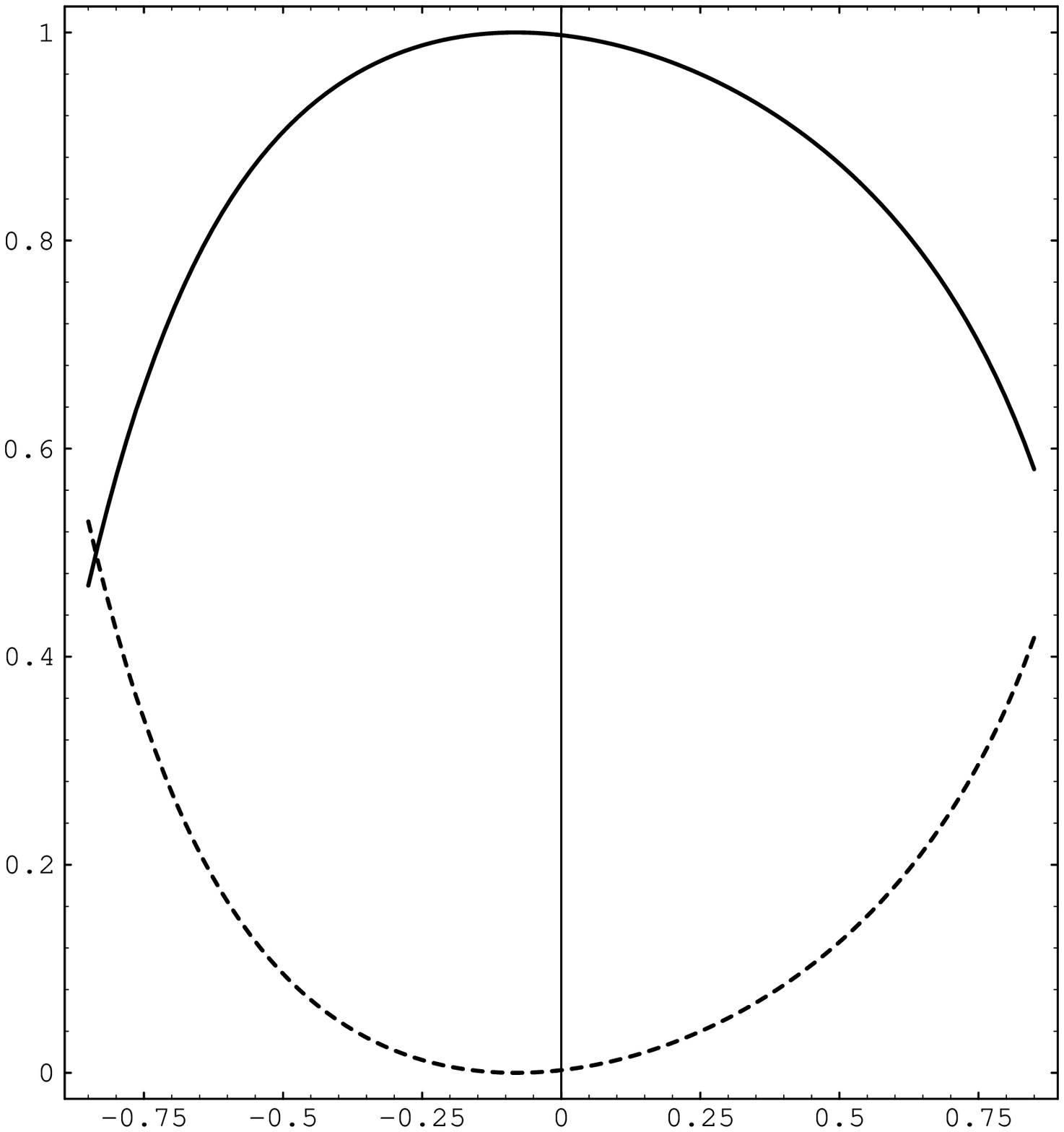,height=22cm}}
 \end{picture}
\end{figure}

\begin{figure}
\vglue -2 true cm
 \unitlength 1mm
 \begin{picture}(175,240)
   \put(17,67){\large $BR(\tilde{t}_1 \to b W \tilde{\chi}^0_1)$}
   \put(17,215){\large $BR(\tilde{t}_1 \to c \tilde{\chi}^0_1)$}
   \put(17,86){\large $BR(\tilde{t}_1 \to c \tilde{\chi}^0_2)$}
   \put(82,52){\large $\cos \theta_t$ }
   \put(66,20){\Large \bf Fig. 6c }
\put(4,32){\psfig{file=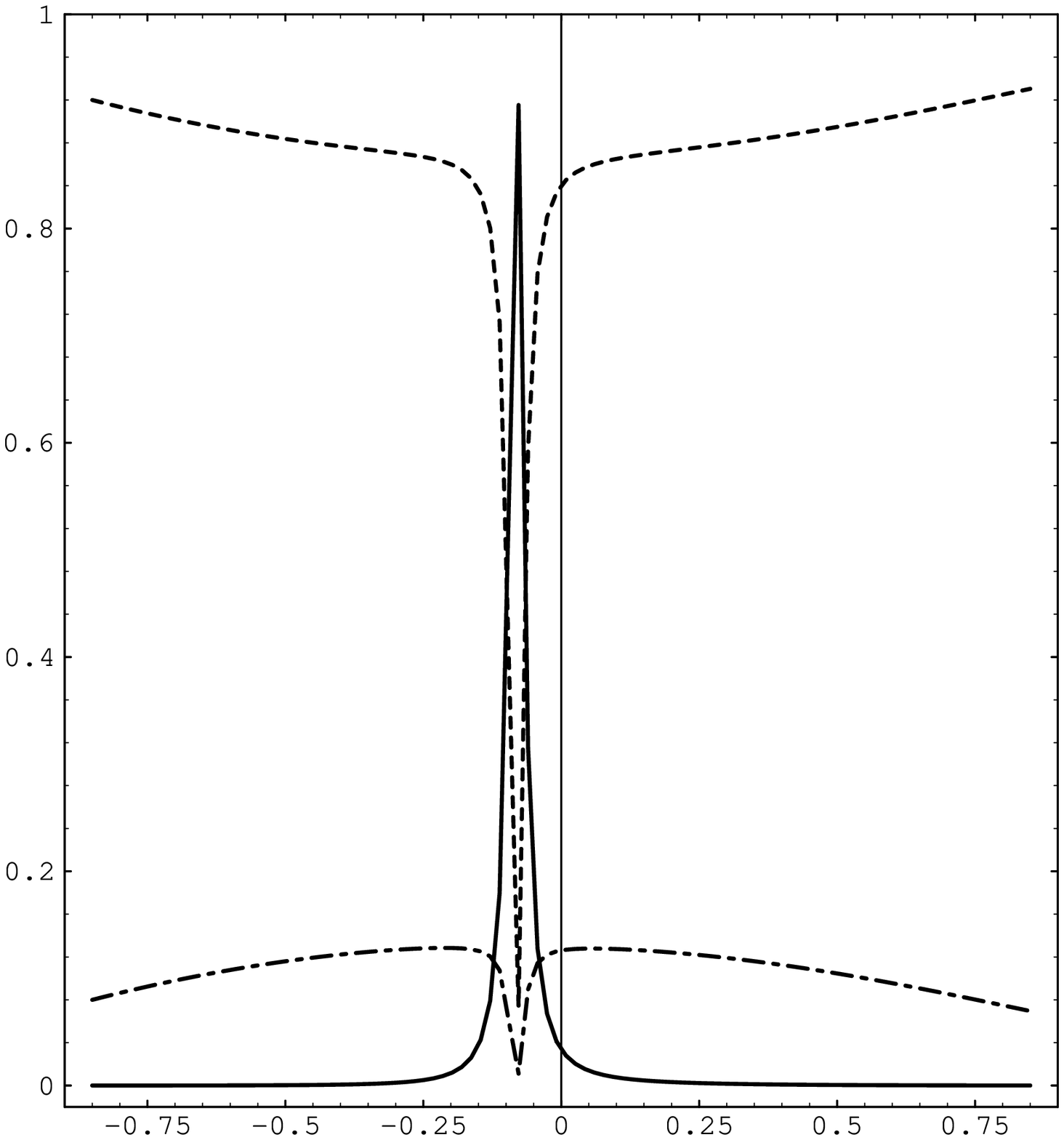,height=22cm}}
 \end{picture}
\end{figure}


\begin{thebibliography}{99}

\bibitem{susy1} For an introduction to supersymmetry, see J.~Wess and
J.~Bagger, {\it Supersymmetry and Supergravity} (Princeton University Press,
1983); P.~Fayet and S.~Ferrara, Phys. Rep. {\bf 32}, 249 (1977); P.~West, {\it
Introduction to Supersymmetry and Supergravity} (World Scientific, 1986);
R.N.~Mohapatra, {\it Unification and Supersymmetry} (Springer-Verlag, 1986).

\bibitem{susy2} For phenomenological reviews of SUSY, see H.P.~Nilles, Phys.
Rep. {\bf 110}, 1 (1984); G.G.~Ross, {\it Grand Unified Theories} (Benjamin
Cummings, 1985); R.~Arnowitt, A.~Chamseddine and P.~Nath, {\it Applied $N=1$
Supergravity} (World Scientific, 1984); H.~Haber and G.~Kane, Phys. Rep. {\bf
117}, 75 (1985); J.~Bagger, lectures at TASI 1995 (to be published)

\bibitem{susy2a}
X.~Tata, lectures at TASI 1995 (to be published), UH-511-833-95,
hep-ph/9510287

\bibitem{Bartl89} A. Bartl, H. Fraas, W. Majerotto and N. Oshimo,
                  Phys. Rev. {\bf D40}, 1594 (1989);
                  A.~Bartl, H. Fraas, W. Majerotto and B. M\"osslacher,
                  Z. Phys. {\bf C55}, 257 (1992).

\bibitem{Ellis83} J. Ellis, S. Rudaz, Phys. Lett. {\bf B218}, 248 (1983);
                  J. F. Gunion, H. E. Haber, Nucl. Phys. {\bf B272}, 1 (1986).

\bibitem{LEPSEARCH}
J.-F. Grivaz, Rapporteur Talk, International Europhysics
Conference on High Energy Physics, Brussels, 1995;
ALEPH collaboration, CERN-PPE/96-10, submitted to Phys. Lett. {\bf B} (1996).
H. Nowak and A. Sopczak, L3 Note 1887, Jan. 1996 ;
S. Asai and S. Komamiya, OPAL Physics Note PN-205, Feb. 1996

\bibitem{hikasa} K. Hikasa and M. Kobayashi, Phys. Rev. {\bf D36}, 724 (1987).

\bibitem{tata3} H. Baer, M. Drees, R. Godbole, and X. Tata, Phys. Rev.
  {\bf D44}, 725 (1991),
  H. Baer, J. Sender, and X. Tata, Phys. Rev. {\bf D50}, 4517 (1994).

\bibitem{D0}
S. Abachi et~al., Phys. Rev. Lett. {\bf 76}, 2222 (1996)

\bibitem{tata1}
A. Bartl, H. Eberl, S. Kraml, M. Majerotto and W. Porod,
UWThPh-1996-35, HEPHY-PUB 646/96, hep-ph/9605412.

\bibitem{denner} A. Denner, H. Eck, O. Hahn, J. K\"ublbeck,
               Nucl. Phys. {\bf B 387}, 467 (1992).

\end{thebibliography}
\end{document}